\documentstyle[twoside,fleqn,espcrc2]{article}

\newcommand{\OO}{{\cal O}}
\newcommand{\bigQ}{\mathrel{\mathop\sim_{Q^2\to\infty}}}

\newcommand{\AmS}{{\protect\the\textfont2
  A\kern-.1667em\lower.5ex\hbox{M}\kern-.125emS}}

\title{Target Independence of the `Proton Spin' Effect}

\author{G.M. Shore \address{Department of Physics,
University of Wales Swansea, \\
Singleton Park, Swansea SA2 8PP, U.K. }
\thanks{Invited talk at QCD 94, Montpellier, July 1994}  }

\begin{document}

\begin{abstract}
Recent work by the author in collaboration with S. Narison and
G. Veneziano on the EMC-SMC-SLAC `proton spin' effect is reviewed.
This uses a novel approach to deep inelastic scattering in which the
matrix elements arising from the OPE are factorised into composite
operator propagators and proper vertices. For polarised $\mu p$
scattering, the composite operator propagator is equated to
the square root of the first moment of the QCD topological
susceptibility, $\sqrt{\chi^\prime(0)}$.
We evaluate $\chi^\prime(0)$ using
QCD spectral sum rules and find a significant suppression relative
to its OZI expectation. This is identified as the source of the
violation of the Ellis-Jaffe sum rule for the first moment of the
polarised proton structure function $g_1^p$.
Our predictions, $\int_0^1 dx g_1^p(x;Q^2=10GeV^2) = 0.143\pm 0.005$
and $\Delta\Sigma =0.353\pm 0.052$, are in excellent agreement with the
new SMC data. This supports our earlier conjecture that the
suppression in the flavour singlet component of the first moment
of $g_1^p$ is a target-independent feature of QCD related to the
$U(1)$ anomaly and is not a special property of the proton structure.

\end{abstract}

\maketitle

\section{Introduction}

The discovery of an anomalous suppression in the first moment of
the polarised proton structure function $g_1^p$
by the EMC collaboration\cite{EMC} in 1988 stimulated a period
of intense theoretical and experimental activity in
the QCD community.
Recently, the existence of this so-called `proton spin
problem' has been confirmed by new experiments at CERN\cite{SMC}
and SLAC\cite{SLAC}, although the numerical results have been
substantially revised.

In this contribution, I will review some recent work\cite{NSV}
with S. Narison and G. Veneziano which we believe gives a
theoretically convincing and quantitative resolution of the problem.
In our picture\cite{SV}, the `proton spin' effect is seen as a natural
addition to the class of OZI-violating $U(1)$ phenomena characteristic
of QCD in the flavour singlet pseudoscalar (or pseudovector) channel.
Furthermore, the violation of the Ellis-Jaffe sum rule, due to
the suppression relative to the OZI expectation of the singlet
form factor $G_A^{(0)}$ (usually denoted by $\Delta\Sigma$),
is recognised as a generic, target-independent
feature of QCD related to the axial $U(1)$ anomaly and not as a special
property of the proton structure. In fact, it reflects an anomalous
suppression of the first moment of the QCD topological susceptibility,
$\chi^\prime(0)$.

\section{The first moment sum rule for $g_1^p$}

The sum rule for the first moment
of the polarised proton structure function $g_1^p$ reads:
\begin{eqnarray}
\Gamma^p_1(Q^2) &&\equiv
\int_0^1 dx g_1^p(x;Q^2) \cr
&&= {1\over 6}  \biggl[ \biggl(G_A^{(3)}(0)
+ {1\over \sqrt3}G_A^{(8)}(0)\biggr)
\biggl(1- {\alpha_s\over\pi}\biggr)\cr
&&~~~~~~~~+ {2\over3} G_A^{(0)}(0;Q^2)
\biggl(1 - {1\over3}{\alpha_s\over\pi}\biggr) \biggr]
\end{eqnarray}
where the $G_A^{(a)}(k^2)$ are form factors in the proton
matrix elements of the axial current:
\begin{equation}
\langle P|J^a_{\mu 5R}(k)|P\rangle =
G_A^{(a)} \bar u \gamma_\mu \gamma_5 u +
G_P^{(a)} k_\mu \bar u \gamma_5 u
\end{equation}
and $a$ is an $SU(3)$ flavour index. Here, we just display the
perturbative corrections to $O\bigl(\alpha_s(Q^2)\bigr)$.
For further terms, see \cite{NSV}.
Since our results depend smoothly on the quark masses in the
chiral limit, we set the light quark masses to zero.

The interpretation of these form factors
in the naive parton model is, in standard notation:
\begin{eqnarray}
G^{(3)}_A(0)&&= {1\over2}(\Delta u-\Delta d) \cr
G^{(8)}_A(0)&&= {1\over{2\sqrt 3}}(\Delta u+\Delta d-2\Delta s)  \cr
G^{(0)}_A(0)&&\equiv \Delta\Sigma = \Delta u+\Delta d+\Delta s
\end{eqnarray}

The axial current occurs here since it is the lowest twist,
lowest spin, odd-parity operator in the OPE of two electromagnetic
currents (see sect.4).
The suffix $R$ emphasises that the current is the renormalised
composite operator. Under renormalisation, the gluon topological
density $Q_R$ and the divergence of the flavour singlet axial
current $J^0_{\mu 5R}$ mix as follows:
\begin{eqnarray}
J^0_{\mu 5R}&&= Z J^0_{\mu 5B} \cr
Q_R&&= Q_B - {1\over2N_F} (1-Z) \partial^\mu J^0_{\mu 5B}
\end{eqnarray}
where $J^0_{\mu 5B} = \sum \bar q \gamma_\mu \gamma_5 q$ and
$Q_B = {\alpha_s\over8\pi} {\rm tr} G^{\mu\nu}\tilde G_{\mu\nu}$
and we have quoted the formulae for $N_F$ flavours.
The mixing is such that the combination occurring in the axial anomaly
Ward identities, e.g.
\begin{equation}
\langle0|\Bigl(\partial^\mu J^0_{\mu 5R} - 2N_F Q_R \Bigr)  \OO
|0\rangle + \langle0|\delta_5 \OO |0\rangle = 0
\end{equation}
is not renormalised.

Since $J^0_{\mu 5R}$ is renormalised, its matrix elements satisfy
renormalisation group equations with an anomalous dimension
$\gamma$, so that in particular $G_A^{(0)}$ depends on the RG scale,
set to $Q^2$ in eq.(1).

We emphasise that $G_A^{(0)}$ does {\it not},
as was initially supposed, measure the spin of the quark constituents
of the proton. The RG non-invariance of $J_{\mu 5R}^0$, which is
a consequence of the anomaly, means that it is {\it not} a
conserved current. Only in the idealised case of a free Dirac field
do the matrix elements of $J_{\mu 5}^0$ and spin coincide.

To compare the sum rule with experiment, we use the following
standard results:
\begin{equation}
G_A^{(3)}(0) = {1\over2}\bigl(F+D\bigr),\hskip0.3cm
G_A^{(8)}(0) = {1\over2\sqrt3}\bigl(3F-D\bigr)
\end{equation}
where $F+D = 1.257\pm0.008$ and $F/D = 0.575\pm0.016$ as fitted from
hyperon and beta decays. We also take
$\alpha_s(m_{\tau}) = 0.347\pm0.030$ from tau decay data.

The Ellis-Jaffe sum rule is obtained by assuming that, in parton
language, the strange quark polarisation in the proton vanishes,
i.e. $\Delta s = 0$ in eq.(3).
This is equivalent to the OZI prediction
\begin{equation}
G_A^{(0)}(0)\big|_{\rm OZI} = 2\sqrt3 G_A^{(8)}(0) = 0.579\pm0.021
\end{equation}
and corresponds to
\begin{equation}
\Gamma_1^p(Q^2=10GeV^2) = 0.170 \pm 0.003
\end{equation}

In contrast, our result
\begin{equation}
G_A^{(0)}(0)\big|_{Q^2=10GeV^2} = 0.353 \pm 0.052
\end{equation}
leads to
\begin{equation}
\Gamma_1^p(Q^2=10GeV^2) = 0.143 \pm 0.005
\end{equation}

We can also compare with the lowest order expectation in the
Skyrme model\cite{EBK}, according to which the proton decouples from
the flavour singlet pseudoscalar channel
($\Gamma_{\Phi_{5R}P \bar P} = 0$
in the language of sect.4) so that $G_A^{(0)}(0) = 0$ and
$\Gamma_1^p(Q^2=10GeV^2) = 0.107 \pm 0.002 $

\begin{figure}[htb]
\vspace{50mm}
\caption{The recent data for $\Gamma_1^p$ from the SMC
collaboration. The data points show
$\int_{x_{\rm min}}^1 dx g_1^p(x)$ at $Q^2 = 10GeV^2$ plotted against
$x_{\rm min}$ and converge
to the SMC value of $\Gamma_1^p(Q^2=10GeV^2) = 0.136\pm0.011\pm0.011$.
Notice that the `world average' quoted by SMC is a little higher
(see eq.(13)).
Also shown are the theoretical predictions of the
Ellis-Jaffe sum rule (EJ), the Skyrme model (S) and our own
prediction (NSV).}
\label{fig:figure 1}
\end{figure}

These theoretical expectations are compared with the experimental
data in Fig.1. The original EMC result was\cite{EMC}
\begin{equation}
\Gamma_1^p(Q^2=11GeV^2) = 0.126 \pm 0.010 \pm 0.015
\end{equation}
which, with the above values for $F/D$ and $\alpha_s$, allows the
following value of $G_A^{(0)}$ to be extracted:
\begin{equation}
G_A^{(0)}\big|_{Q^2=11GeV^2} = 0.19 \pm 0.17
\end{equation}

Clearly, this value is barely consistent with our prediction.
It is therefore extremely gratifying that the improved analysis by the
SMC collaboration quoted in \cite{SMC} now gives the new `world
average'
\begin{equation}
\Gamma_1^p(Q^2=10GeV^2) = 0.145 \pm 0.008 \pm 0.011
\end{equation}
from which we deduce
\begin{equation}
G_A^{(0)}\big|_{Q^2=10GeV^2} = 0.37 \pm 0.07 \pm 0.10
\end{equation}
Notwithstanding the large experimental errors, the agreement
between the new data and our prediction is excellent.

\section{The Composite Operator/Proper Vertex Method for DIS}

The essential features of our method are easily described for a
general deep inelastic scattering process. The hadronic part of the
scattering amplitude is given by the imaginary part of the
two-current matrix element $\langle N| J_\mu(q) J_\nu(-q) |N\rangle$.
The OPE is used to expand the large $Q^2$ limit
of the product of currents as a sum of Wilson coefficients
$C_i(Q^2)$ times renormalised composite operators $\OO_i$ as
follows (suppressing Lorentz indices),
\begin{equation}
J(q) J(-q)  \bigQ   \sum_i C_i(Q^2) \OO_i(0)
\end{equation}
The dominant contributions to the amplitude arise from the operators
$\OO_i$ of lowest twist. Within this set of lowest twist operators,
those of spin $n$ contribute to the $n$th moment of the structure
functions, i.e.
\begin{equation}
\int_0^1 dx  x^{n-1} F(x;Q^2) = \sum_i C_i^n(Q^2) \langle N|
\OO_i^n(0) |N\rangle
\end{equation}

The Wilson coefficients are calculable in QCD perturbation theory,
so the problem reduces to evaluating the target matrix elements of the
corresponding operators. We now introduce appropriately defined
proper vertices $\Gamma_{\tilde\OO N\bar N}$, which are chosen to be 1PI
with respect to a physically motivated basis set $\tilde\OO_j$
of renormalised composite operators. The matrix elements are then
decomposed into products of these vertices with zero-momentum
composite operator propagators as follows,
\begin{equation}
\langle N| \OO_i(0) |N\rangle =  \sum_j \langle 0|\OO_i(0)
\tilde\OO_j(0) |0\rangle~ \Gamma_{\tilde\OO_j N\bar N}
\end{equation}
Despite being non-perturbative, we can frequently evaluate the
composite operator propagators using a combination of exact
Ward identities and dynamical approximations.


All this is illustrated in \cite{NSV}.
In essence, what we have done is to
split the whole amplitude into the product of a `hot QCD' (high
momentum) part described by QCD perturbation theory, a `cold QCD'
part described by a (non-perturbative, zero-momentum) composite operator
propagator and finally a target-dependent proper vertex.
The generic expression for a structure function sum rule is then:
\begin{eqnarray}
&&\int_0^1 dx  x^{n-1} F(x;Q^2)  \cr
&&= \sum_i \sum_j C_i^n(Q^2)~
\langle 0|O_i(0) \tilde\OO_j(0) |0\rangle~
\Gamma_{\tilde\OO_j N\bar N}~~~~~~
\end{eqnarray}

All the target dependence is contained in the vertex function
$\Gamma_{\tilde\OO N\bar N}$.
However, these are not unique -- they depend
on the choice of the basis $\tilde\OO_j$ of composite operators.
This choice is made on physical grounds based on the relevant
degrees of freedom, the aim being to parametrise the amplitude
in terms of a minimal, but sufficient, set of vertex functions.
(These play the r\^ole of the non-perturbative parton
distributions in the usual treatment).
A good choice can often lead to an almost direct correspondence
between the proper vertices and physical couplings such as, e.g.,
the pion-nucleon coupling $g_{\pi NN}$. In particular, the
proper vertices should be chosen whenever possible to be RG invariant.

It is important to realise that the decomposition (17) is an
{\it exact} expression, independent of the choice of the set of
operators $\tilde\OO_j$. A different choice of basis set
merely changes the definition of the proper vertices.
In particular, it is {\it not} to be understood that the set
$\tilde\OO_j$ is in any sense a complete set and that choosing
a finite number of operators (such as the pseudoscalars
$\Phi_{5R}$ and $Q_R$ in sect.4) represents an approximation.

\section{The $g_1^p$ Sum Rule and the Topological Susceptibility}

We now apply this method to the sum rule for $g_1^p$. The relevant
OPE is:
\begin{equation}
J_\mu(q) J_\nu(-q)   \bigQ  2\sum_{a=0,3,8}
\epsilon_{\mu\nu\alpha}{}^\beta
{q^\alpha \over Q^2} C^a(Q^2) J^a_{\beta 5R}
\end{equation}
Assuming the absence of a massless pseudoscalar (Goldstone) boson
in the $U(1)$ channel, we find
\begin{eqnarray}
G_A^{(0)}(0;Q^2) \bar u \gamma_5 u&&=
{1\over 2M} \langle P|\partial^\mu J^0_{\mu 5R} |P\rangle \cr
&&= {1\over 2M} 2N_F \langle P| Q_R(0) |P\rangle
\end{eqnarray}
where we have used the anomalous chiral Ward identity
to re-express $G_A^{(0)}(0)$ as the forward matrix element
of the renormalised gluon topological density $Q_R$.
$M$ is the proton mass.

We now choose the composite operator basis $\tilde\OO_j$ to be the
set of renormalised flavour singlet pseudoscalar operators
$Q_R$ and $\Phi_{5R}$, where,
up to a subtle but crucial normalisation factor (see \cite{NSV}
and \cite{SV} for an explanation), the corresponding bare operator
is simply $i\sum\bar q \gamma_5 q$.
We may then write (c.f. eq.(17)):
\begin{eqnarray}
&&\langle P|Q_R(0)|P\rangle  \cr
&&= \langle 0|Q_R Q_R|0\rangle \Gamma_{Q_R P \bar P} +
\langle 0|Q_R \Phi_{5R}|0\rangle \Gamma_{\Phi_{5R} P \bar P}~~~~~~
\end{eqnarray}
where the composite operator propagators are at zero momentum
and the proper vertices are 1PI with respect to $Q_R$
and $\Phi_{5R}$ only.

The composite operator propagator in the first term in eq.(21)
is the zero-momentum limit of an important quantity in QCD known
as the topological susceptibility $\chi(k^2)$, viz.
\begin{equation}
\chi(k^2) = \int dx e^{ik.x} i\langle 0|T^* Q_R(x)~Q_R(0)|0\rangle
\end{equation}
Moreover, it can be shown exactly using chiral Ward identities\cite{NSV}
that the propagator $\langle 0|Q_R~\Phi_{5R}|0\rangle$ at zero momentum
is simply the square root of the first moment of the topological
susceptibility. We therefore find:
\begin{equation}
\langle P| Q_R(0) |P\rangle = \chi(0) \Gamma_{Q_R P \bar P}
+\sqrt{\chi^{\prime}(0)} \Gamma_{\Phi_{5R} P \bar P}
\end{equation}

The chiral Ward identities also show that for QCD with massless
quarks, $\chi(0)$ actually vanishes. This is in contrast to pure
Yang-Mills theory, where $\chi(0)$ is non-zero.
We therefore arrive at our basic result\cite{SV}:
\begin{equation}
G_A^{(0)}(0;Q^2) \bar u \gamma_5 u
= {1\over 2M} 2N_F \sqrt{\chi^{\prime}(0;Q^2)}
\Gamma_{\Phi_{5R} P \bar P}
\end{equation}

The quantity $\sqrt{\chi^\prime(0)}$ is
not RG invariant and scales with the anomalous dimension $\gamma$.
On the other hand, the proper vertex has been chosen specifically
so as to be RG invariant. The renormalisation group properties of this
decomposition are crucial to our resolution of the `proton spin problem'.

Our proposal is that we should expect the source of OZI violations
to lie in RG non-invariant terms, i.e. in $\chi^{\prime}(0)$.
The reasoning is as follows. In the absence of the $U(1)$
anomaly, the OZI rule would be an exact property of QCD. So the OZI
violation is a consequence of the anomaly. But it is the existence
of the anomaly that is responsible for the non-conservation and
hence non-trivial renormalisation of the axial current $J^0_{\mu 5R}$.
We therefore expect to find OZI violations in quantities sensitive
to the anomaly, which we identify through their RG dependence
on the anomalous dimension $\gamma$.
This seems reasonable since, if the OZI rule were to be good for
such quantities, it would mean approximating a RG non-invariant,
scale-dependent quantity by a scale-independent one.

Notice that we are {\it not} saying that the OZI violation is due
to a large (non-perturbative) scaling effect dependent on $\gamma$.
We are simply using the dependence on the anomalous dimension to
identify those quantities most likely to display significant
differences from their OZI approximations.

If this proposal is correct, we expect $\sqrt{\chi^{\prime}(0)}$ to be
significantly suppressed relative to its OZI approximation
of $(1/\sqrt{6}) f_\pi$.
The proper vertex $\Gamma_{\Phi_{5R} P \bar P}$
would behave exactly as expected according to the OZI rule.
That is, the Ellis-Jaffe violating suppression of the first moment
of $g_1^p$ observed by EMC would {\it not} be a special property of the
proton at all, but would simply be due to an anomalously small
value of the first moment of the QCD topological susceptibility
$\chi^{\prime}(0)$.

This is our conjectured resolution\cite{SV} of the
`proton spin problem'. It is further
supported by a number of experimental results in the pseudoscalar
$U(1)$ channel, notably in $\eta^\prime \rightarrow \gamma\gamma$
decays. See \cite{SV} for further discussion.

Putting all this together, we conjecture the following expression
for the singlet form factor:
\begin{equation}
G_A^{(0)}(0;Q^2) = 2\sqrt3 G_A^{(8)}(0)~
{\sqrt{\chi^\prime(0;Q^2)}\over
\bigl(f_\pi /\sqrt6\bigr) }
\end{equation}

\section{$\chi^\prime(0)$ from QCD Spectral Sum Rules}

The remaining task is to find a non-perturbative estimate of the
first moment of the topological susceptibility, $\chi^\prime(0)$.
This is a fundamental quantity likely to reappear in
many applications of QCD, and an evaluation from first principles
represents a strong challenge to lattice gauge theory. Of course,
for a meaningful result it is necessary to work beyond the
quenched approximation, close to the chiral limit.

Instead, we estimate $\chi^\prime(0)$ using QCD spectral sum rules.
A full description of the calculation is given in \cite{NSV} and
here we only quote the result. We have evaluated $\chi^\prime(0)$
using subtracted dispersion relations with the Laplace sum rule
method, finding good stability, and have confirmed the result using
the finite energy sum rule technique. The spectral function is
saturated with the single lightest pseudoscalar state,
the $\eta^\prime$. We find
\begin{equation}
\sqrt{\chi^\prime(0)}\Big|_{Q^2=10 GeV^2}
= 23.2 \pm 2.4~{\rm MeV}~,
\end{equation}
a suppression of approx.~$0.6$ relative to the OZI value
$f_\pi /\sqrt6$. Substituting this into eq.(25)
finally gives our result (9) for $G_A^{(0)}$.

The essential input parameter in the spectral sum rules is the
$\eta^\prime$ mass. (A quite different result is found for
$\chi^\prime(0)$ in pure Yang-Mills theory, saturating the
spectral function with a pseudoscalar glueball.)
In essence, the sum rules allow us to determine the relevant mass
scale for OZI breaking in the pseudovector channel using as
input the known OZI-violating $\eta^\prime$ mass from the
pseudoscalar channel. The link is the $U(1)$ Goldberger-Treiman
relation\cite{SV} which underlies our approach.

\section{Conclusion: Not Spin, Not the Proton, Not a Problem!}

Our conclusions are simply stated. $\Delta\Sigma$ does {\it not}
measure spin. Its suppression relative to the OZI (Ellis-Jaffe)
value is due to an anomalously small value of $\chi^\prime(0)$
in QCD and is {\it not} a special property of the proton.
The violation of the Ellis-Jaffe sum rule is {\it not} a problem
in QCD -- the flavour singlet pseudovector channel is precisely
where we should expect to find large OZI violations and, using
spectral sum rules, we have given a successful quantitative
prediction of $\Delta\Sigma$ and $\Gamma_1^p$.

The $g_1^p$ sum rule does, however, present some problems
for QCD-inspired models of the proton. The Skyrme model would have
to be significantly extended to incorporate the $O(1/N_C)$ effects
characteristic of the $U(1)$ anomaly. In the parton model,
the effect can be incorporated (though not as yet predicted)
by modifying the constituent quark expression (3) for $\Delta\Sigma$
to include a polarised gluon density $\Delta\Gamma$ with
the necessary renormalisation group behaviour.
For further details of this approach, see \cite{AR}.

Finally, it would be interesting to test our proposal of
target independence directly by experiment. This should be possible
in semi-inclusive processes in which a pion or D meson is detected.


\begin{thebibliography}{9}
\bibitem{EMC} EMC: J. Ashman et al., Phys. Lett. B206 (1988) 364;
Nucl. Phys. B328 (1990) 1.
\bibitem{SMC} SMC: D. Adams et al., Phys. Lett. B329 (1994) 399.
\bibitem{SLAC} E143: V. Breton, these proceedings.
\bibitem{NSV} S. Narison, G.M. Shore and G. Veneziano, CERN-TH.7223/94
\bibitem{SV} G.M. Shore and G. Veneziano, Nucl. Phys. B381 (1992) 23
\bibitem{EBK} J. Ellis, S.J. Brodsky and M. Karliner, Phys. Lett.
B206 (1988) 309.
\bibitem{AR} G. Altarelli and G. Ross, Phys. Lett. B212 (1988) 391;
G. Altarelli, these proceedings.
\end{thebibliography}
\end{document}